\documentclass[aps,prl,preprint,superscriptaddress,showpacs]{revtex4}

\usepackage{textcomp}
\usepackage{amsmath}
\usepackage{graphicx}

\begin{document}

\title{Ground-State Structures of Ice at High-Pressures}

\author{Jeffrey M. McMahon}
\email[]{mcmahonj@illinois.edu}
\affiliation{Department of Physics, University of Illinois at Urbana-Champaign, Illinois 61801, USA}

\date{\today}

\begin{abstract}
\textit{Ab initio} random structure searching based on density functional theory is used to determine the ground-state structures of ice at high pressures. Including estimates of lattice zero-point energies, ice is found to adopt three novel crystal phases. The underlying sub-lattice of O atoms remains similar among them, and the transitions can be characterized by reorganizations of the hydrogen bonds. The symmetric hydrogen bonds of ice X and $Pbcm$ are initially lost as ice transforms to structures with symmetries $Pmc2_1$ ($800$ -- $950$ GPa) and $P2_1$ ($1.17$ TPa), but they are eventually regained at $5.62$ TPa in a layered structure $C2/m$. The $P2_1 \rightarrow C2/m$ transformation also marks the insulator-to-metal transition in ice, which occurs at a significantly higher pressure than recently predicted.
\end{abstract}



\maketitle



The behavior of $\text{H}_2\text{O}$ at high pressures is of fundamental important for both condensed matter and planetary physics \cite{Hobbs_IcePhys, Hubbard_PlanetaryInteriors}. This can be attributed to its substantial abundance in the universe, and the fact that a significant fraction of it exists in ice form at high pressures in planetary interiors. In our solar system alone, for example, Uranus and Neptune consist largely of $\text{H}_2\text{O}$, ammonia, and methane ice mixtures up to $800$ GPa, and the cores of Saturn and Jupiter likely contain ice components at pressures of approximately $800$ GPa -- $1.8$ TPa and $4$ -- $5$ TPa, respectively \cite{planet-interiors_Guillot-Science-1999}. Despite this importance, very little is known about the behavior of solid $\text{H}_2\text{O}$ (ice) under such extreme conditions. The primary reasons for this are that experiments have thus far only reached $210$ GPa \cite{H2O_exp_210GPa_Hemley-Science-1996} and, until recently, theoretical and computational methods have not existed to reliably predict crystal structures with little or no a priori information.

The known phase diagram of $\text{H}_2\text{O}$ is extremely rich. $15$ thermodynamically stable phases of ice have been observed experimentally \cite{H2O-structs_review_Malenkov-JPhysCondMat-2009}, ice XV only recently \cite{H2O_ice-XV_Finney-PRL-2009}. The highest pressure phase experimentally observed is ice X, which is obtained from a phase-transition from ice VII (or VIII, depending on the temperature) near $44$ GPa \cite{H2O_ice-X_Grimsditch-PRL-1984}. In this phase, the $\text{O}$ atoms form a body-centered cubic sub-lattice and the $\text{H}$ atoms adopt symmetric positions between them at pressures near $110$ -- $120$ GPa \cite{H2O_H-bond_centering_Benoit-PRL-2002}, and so this phase is also referred to as symmetric ice. Because of this the distinction between covalent bonds and hydrogen bonds is lost, as is thus the molecular form of $\text{H}_2\text{O}$, resulting in an atomic solid. Recent lattice dynamical calculations using density functional theory (DFT) suggest that the symmetric ordered form of ice X is only stable from $120$ -- $400$ GPa \cite{H2O_ice-X_instabilities_Caracas-PRL-2008}.
Near $300$ -- $400$ GPa a lattice instability occurs, resulting in a transition to a crystal phase with $Pbcm$ symmetry (Hermann--Mauguin space-group symbol) \cite{H2O_ice-X_instabilities_Caracas-PRL-2008}, as first predicted by Benoit \textit{et al.} in $1996$ via constant pressure molecular dynamics simulations \cite{H2O_Pbcm_Parrinello-PRL-1996}. This phase results from a compromise between the packing efficiency of the O atoms and a preservation of the symmetric hydrogen bonds, resulting in a distorted hexagonal close-packed (hcp) sub-lattice of O atoms.
At even higher pressures, which is of fundamental importance to planetary physics, for example, very little is known. However, a number of intriguing possibilities have been proposed, perhaps the most interesting being an insulator-to-metal transition \cite{H2O_metallization_Grimsditch-BookCh-1985, H2O_Cmcm_Pbca_Militzer-PRL-2010}.


In this Letter, the recently proposed \textit{ab initio} random structure searching (AIRSS) method of Pickard and Needs to predict crystal structures \cite{AIRSS_orig_Pickard-PRL-2006}, combined with DFT, is used to determine the ground-state structures of ice at high pressures. In this method, a number of random configurations are each relaxed to the ground state at constant pressure (see below). After enough trials, a good sampling of the configuration space is obtained and the ground-state structure(s) can be identified. This method has been used to successfully predict the ground-state structures of a number of systems \cite{AIRSS_review_Pickard-JPhysCondMatt-2011}, most recently including those of atomic metallic hydrogen \cite{McMahon_atomic-H_ground-state_2011}.

Calculations were performed using the Quantum ESPRESSO DFT code \cite{QE-2009}. Norm-conserving Troullier--Martins pseudopotentials \cite{TM-PP_Troullier-Martins-PRB-1991} were used for all calculations. For O, core radii of $1.25$, $1.25$, and $1.4$ a.u.\ were used for the $s$, $p$, and $d$ components, respectively. For H, a core radius of $0.8$ a.u.\ was used for the AIRSS and then decreased to $0.3$ a.u.\ for recalculating detailed enthalpy vs pressure curves. For all calculations, the Perdew-Burke-Ernzerhof generalized-gradient approximation exchange and correlation functional \cite{PBE_exch-correl_PRL-1996} was used. See Ref.\ \cite{SI_ref} for justifications for these approximations. A plane-wave basis set with a cutoff of $120$ Ry was used for the AIRSS and then increased to $350$ Ry for recalculating enthalpy curves. For Brillouin-zone sampling, $8^3$ \textbf{k}-points were used for all calculations, except for $Cmcm$, $Cmca$, $P4_2/nnm$, and $P2_1/m$ (see below) for which $12^3$ were used to recalculate enthalpy curves. The high-accuracy cutoff and \textbf{k}-point sampling were found to give a total convergence in energy to better than $1$ mRy/$\text{H}_2\text{O}$ and the total pressure to approximately $1$ GPa for each structure.
Phonons were calculated using density functional perturbation theory as implemented within Quantum ESPRESSO, and were converged to a similar level of accuracy as the DFT calculations.


Random structures were constructed by generating random unit-cell translation vectors, renormalizing the volume, and choosing random $\text{H}_2\text{O}$ configurations (positions and orientations). Constant pressure geometry relaxations were performed at $0.5$, $1$, $1.5$, $2$, $3$, and $5$ TPa for unit cells containing $4$ $\text{H}_2\text{O}$ units, and then additional relaxations were performed at $1$ and $2$ TPa using unit cells containing $6$ and $8$ $\text{H}_2\text{O}$ units. (However, the latter searches only revealed a couple of additional structures -- see below and Ref.\ \cite{SI_ref}.) It is important to realize that searches over unit cells of their factors are implicitly included in these calculations -- i.e., those with $1$, $2$, or $3$ $\text{H}_2\text{O}$ units. While structures with unit cells containing $5$, $7$, or more $\text{H}_2\text{O}$ units are certainly possible, it is reasonable to suspect that they are unlikely based on comparisons with the other predicted high-pressure phases of ice, such as $Pbcm$ \cite{H2O_Pbcm_Parrinello-PRL-1996} and the recently proposed $Cmcm$ and $Pbca$ structures \cite{H2O_Cmcm_Pbca_Militzer-PRL-2010} (all found via simulations capable of generating unit cells with up to $16$ $\text{H}_2\text{O}$ units). Typical relaxations included up to $175$ random structures at each pressure considered, which appeared to be enough to generate both the lowest-enthalpy structure and higher-enthalpy metastable ones multiple times. 

As a confirmation of the validity of this method to reliably predict the high-pressure phases of ice, at $500$ GPa $10\%$ of the relaxations revealed $Pbcm$ as the lowest-enthalpy structure. This is consistent with previous predictions \cite{H2O_Pbcm_Parrinello-PRL-1996, H2O_ice-X_instabilities_Caracas-PRL-2008}, indicating that AIRSS as described above is indeed capable of generating (presumably correct) ground-state structures with a high probability and that the searches are exhaustive, as well as supports the ice X $\rightarrow$ $Pbcm$ transformation.

\begin{figure}
  \includegraphics[scale=0.28, bb=0 0 875 657]{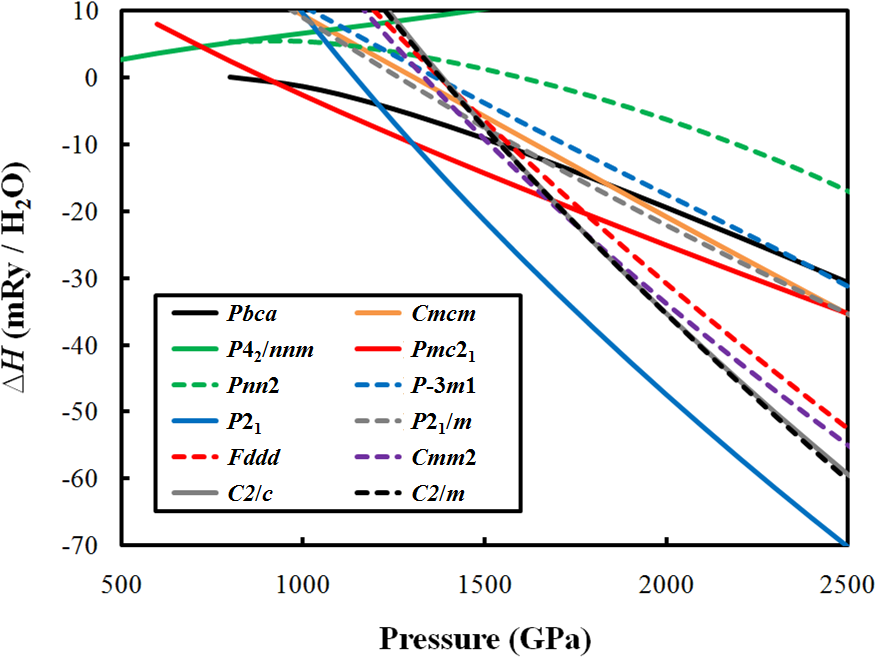}
  \caption{(color online). Enthalpies of the ground-state structures of ice relative to $Pbcm$, not including lattice zero-point energies. Note that the enthalpies vs pressure are nearly linear from $2.5$ -- $5$ TPa (not shown).}
  \label{fig:phase-diag}
\end{figure}

After performing the AIRSS, each structure within approximately $25$ mRy/$\text{H}_2\text{O}$ of the lowest-enthalpy one found was considered for further investigation. Detailed enthalpy vs pressure curves were calculated for these structures by performing additional constant-pressure geometry optimizations while keeping symmetries fixed; Fig.\ \ref{fig:phase-diag}.
%
%

\begin{figure}
  \includegraphics[scale=0.34, bb=0 0 718 350]{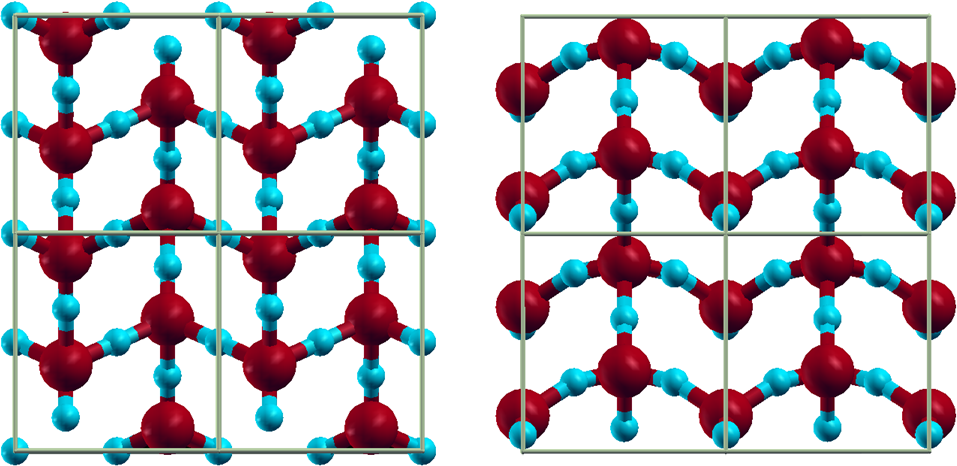}
  \caption{(color online). Ground-state structures of ice at $1$ TPa. (left) $Pbcm$ and (right) $Pmc2_1$.}
  \label{fig:Pbcm_Pmc21}
\end{figure}

Near $800$ GPa, $Pbcm$ becomes unstable relative to two additional structures, $Pmc2_1$ and $Pbca$. Such instability is expected, as it has been demonstrated that $Pbcm$ develops a dynamic instability near $760$ GPa in the ($1/2$, $0$, $0$) phonon mode \cite{H2O_Cmcm_Pbca_Militzer-PRL-2010}. $Pbcm$ and $Pmc2_1$ are both shown in Fig.\ \ref{fig:Pbcm_Pmc21}, and $Pbca$ is shown in Refs.\ \cite{SI_ref, H2O_Cmcm_Pbca_Militzer-PRL-2010}. 
$Pmc2_1$ and $Pbca$ are both similar to $Pbcm$. For example, in the perspective of Fig.\ \ref{fig:Pbcm_Pmc21} the O atoms remain close to their distorted hcp sub-lattice positions \cite{H2O_Pbcm_Parrinello-PRL-1996}. However, the H atoms are shifted away from their symmetric O--H--O positions. In $Pbca$, a small distortion of the H atoms occurs in alternating directions, leaving them close to tetrahedral sites and the hydrogen-bond network intact. In $Pmc2_1$, on the other hand, a reorganization of the hydrogen-bond network occurs. In the perspective of Fig.\ \ref{fig:Pbcm_Pmc21}, every O atom in $Pbcm$ is connected to both its vertical and horizontal neighbors via symmetric hydrogen bonds. In $Pmc2_1$, this bonding is only retained in every other column of O atoms, but without symmetric hydrogen bonds. The other O atoms become disconnected from their vertical neighbors, and instead become hydrogen bonded (and not symmetrically) with O atoms out of the plane.

It should be noted that $Pbca$ was recently proposed as a likely candidate for the ground-state structure of ice from $760$ GPa to $1.25$ TPa \cite{H2O_Cmcm_Pbca_Militzer-PRL-2010}. However, Fig.\ \ref{fig:phase-diag} shows that it is only competitively stable with $Pmc2_1$ from $800$ -- $925$ GPa. It is certainly possible that $Pbca$ is a stable phase of ice in this narrow pressure range. However, given that it is only a slight distortion of $Pbcm$ and less than $1$ mRy/$\text{H}_2\text{O}$ more stable, a more likely scenario is that it is a transitional structure that occurs during the $Pbcm \rightarrow Pmc2_1$ transformation (this is further suggested by lattice dynamics calculations that show $Pbca$ is unstable near $800$ GPa -- see Ref.\ \cite{SI_ref}). In any case, $800$ GPa (approximately) marks a transition away from the symmetric hydrogen bonds seen in ice X and $Pbcm$, which are shown below to be regained at much higher pressures.

\begin{figure}
  \includegraphics[scale=0.26, bb=0 0 920 471]{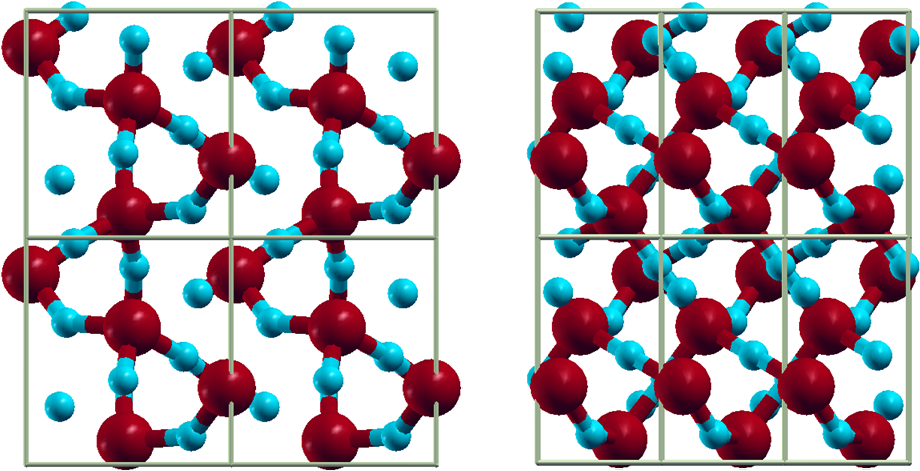}
  \caption{(color online). $P2_1$ at $1.4$ TPa. (left) Same perspective as in Fig.\ \ref{fig:Pbcm_Pmc21}. (right) Side view.}
  \label{fig:P21}
\end{figure}

$Pmc2_1$ remains stable until $1.3$ TPa. It then becomes relatively unstable towards a structure with $P2_1$ symmetry, which is shown in Fig.\ \ref{fig:P21}.
Comparison of Figs.\ \ref{fig:Pbcm_Pmc21} and \ref{fig:P21}(left) shows that (in the plane of each figure) the O atoms continue to remain close to their distorted hcp positions. Into the plane, however, $P2_1$ undergoes a noticeable compression and slight distortion relative to $Pbcm$ (not shown). Moreover, it can be seen that a further reorganization of the hydrogen-bond network occurs. O atoms in every other column continues to remain connected to their vertical neighbors, but the hydrogen bonds distort slightly outwards in alternating directions. The hydrogen bonds of the other O atoms, on the other hand, rearrange more significantly, and each O atom becomes connected to either one neighboring column or the other, also in an alternating fashion.

\begin{figure}
  \includegraphics[scale=0.21, bb=0 0 1149 439]{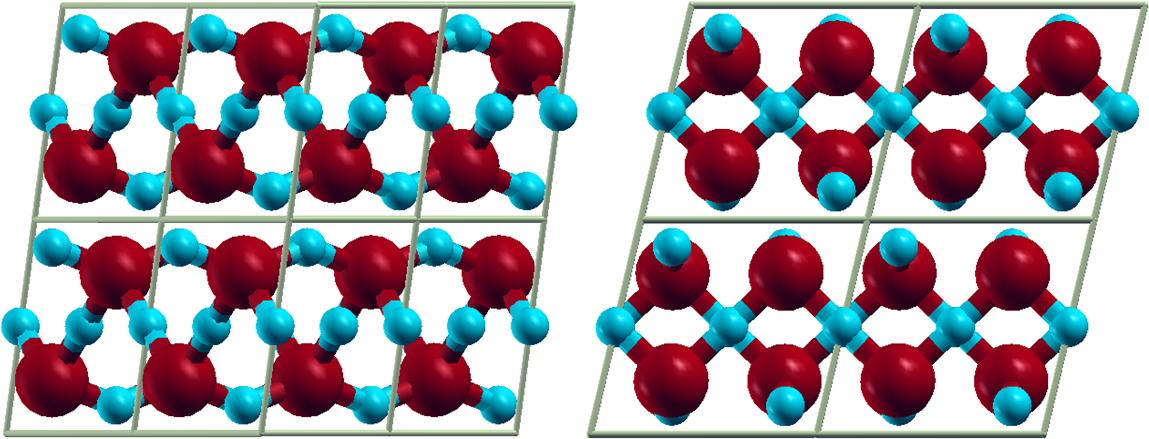}
  \caption{(color online). $C2/m$ at $5$ TPa. The perspectives are similar to those shown in Fig.\ \ref{fig:P21}. Note that alternating O atoms are out of the plane relative to one another.}
  \label{fig:C2m_C2c}
\end{figure}

$P2_1$ remains the lowest-enthalpy structure up to the highest pressure considered in this work of $5$ TPa. However, another competitive structure was also found in this pressure range, $C2/m$. In fact, Fig.\ \ref{fig:phase-diag} shows that the relative enthalpy difference between $C2/m$ and $P2_1$ slowly decreases with increasing pressure. A linear extrapolation of the enthalpy vs pressure curves near $5$ TPa (which should be quite accurate) indicates a transition pressure of approximately $5.29$ TPa. As can be seen in Fig.\ \ref{fig:C2m_C2c}, $C2/m$ is a layered structure, where each layer consists of two sets of O atoms.
Additionally, a slight shear deformation of the layers occurs, leaving the O atoms close to, but slightly displaced from their distorted hcp positions. (Note that a less stable structure without the shear deformation, $C2/c$, was also found -- see Ref.\ \cite{SI_ref}.) Moreover, symmetric hydrogen bonds are seen to be regained, which connect the O atoms within each layer. 

Comparison of the O atoms in Fig.\ \ref{fig:P21}(right) with \ref{fig:C2m_C2c}(left) indicates that a transition towards the layered $C2/m$ structure is evident in $P2_1$ (and in fact even $Pbcm$, which looks similar but less compressed, as discussed above). Given these results, it is possible to understand the phase transformations that ice undergoes at high pressures. In all structures, the O atoms remain close to their distorted hcp positions and compress along a preferred axis (e.g., into the plane of Fig.\ \ref{fig:Pbcm_Pmc21}). This compression leads to structures without symmetric hydrogen bonds, $Pmc2_1$ and $P2_1$, obtained via reorganizations of the hydrogen-bond network. Continued compression eventually results in a layered structure where the symmetric hydrogen bonds are regained.

\begin{figure}
  \includegraphics[scale=0.17, bb=0 0 1383 609]{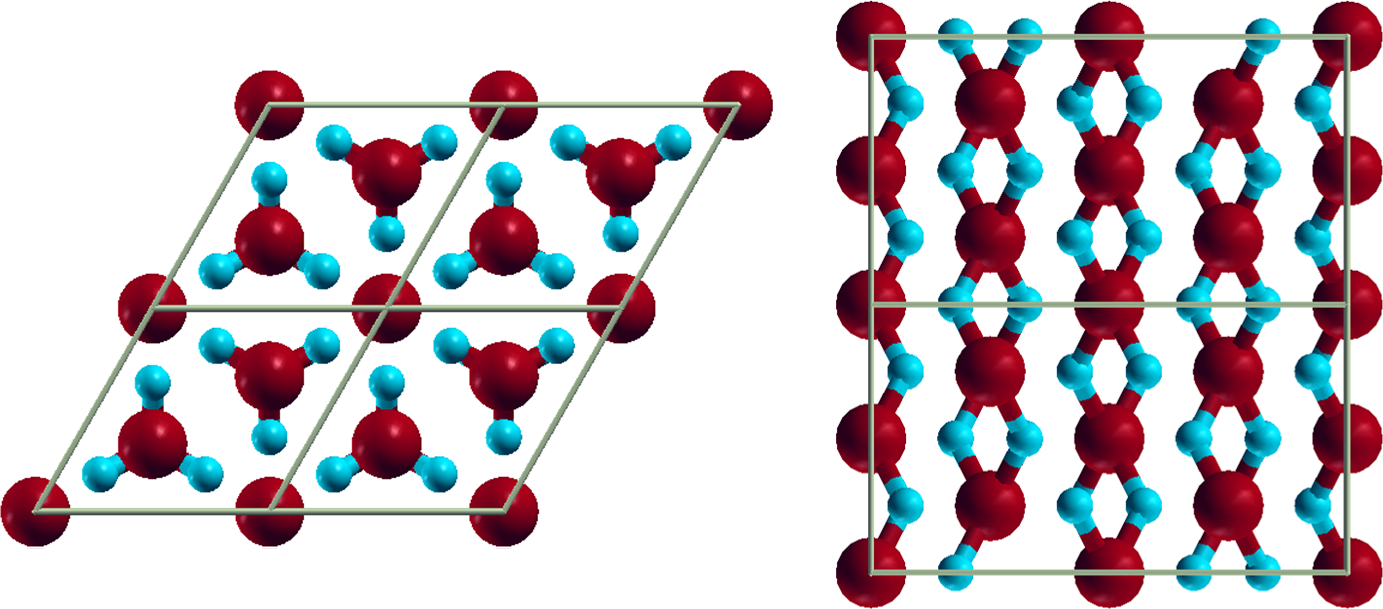}
  \caption{(color online). Metastable structures of ice at $2$ TPa. (left) $P$-$3m1$ and (right) $Fddd$.}
  \label{fig:metastable_structs}
\end{figure}

A number of structures with higher enthalpies were also generated during the AIRSS, as evident from Fig.\ \ref{fig:phase-diag}. It has recently been demonstrated (at medium pressure) that such metastable structures of ice are both experimentally realizable \cite{H2O_metastable_Hemley-Science-2006} and predictable via AIRSS \cite{AIRSS_metastable-H2O_Pickard-JCP-2007}. It is thus quite possible that those shown in Fig.\ \ref{fig:phase-diag} may actually occur in nature, and therefore worthwhile to very briefly mention a couple of them here. Brief descriptions of the other ones can be found in Ref.\ \cite{SI_ref}. Two of the most interesting metastable structures found were $P$-$3m1$ and $Fddd$, as shown in Fig.\ \ref{fig:metastable_structs}.
In both structures, the O atoms once again remain near distorted hcp positions. However, their hydrogen bonds are quite different. In $P$-$3m1$, for example, $2/3$ of the O atoms have $3$ strongly-coordinated H atoms and the other $1/3$ have none, but $6$ weakly-coordinated ones instead. 
$Fddd$, on the other hand, forms a string-like structure though its hydrogen bonds. (Note that below $1.6$ TPa, $Fddd$ distorts into a lower symmetric form, $Fdd2$.)


The results presented above were for static lattices. However, the light hydrogen mass causes the phases of ice at high pressures to have large zero-point energies (ZPEs) that must be estimated in order to determine the most stable ground-state structures. ZPEs were neglected during the AIRSS, but their impacts were estimated afterwards using the harmonic approximation: $E_\text{ZPE} = \int d\omega ~ F(\omega) \hbar \omega / 2$, where $F(\omega)$ is the phonon density of states. $F(\omega)$ was calculated using a $2^3$ grid of \textbf{q}-points in the Brillouin zone, which is estimated to be sufficient to converge ZPE differences between structures to within a few percent.

Reference \cite{SI_ref} shows that the ZPEs are quite large, increasing from approximately $67$ mRy/$\text{H}_2\text{O}$ at $400$ GPa to $127$ mRy/$\text{H}_2\text{O}$ at $5$ TPa. Despite such large values, ZPE differences between the structures are relatively small, in all cases within a few mRy/$\text{H}_2\text{O}$.
While this energy scale is not enough to change the ordering of the structures, it is enough to affect precise transition pressures, in some cases. For example, $P2_1$ is found to have a lower ZPE than both $Pmc2_1$ and $C2/m$, causing the corresponding transition pressures to be shifted to $1.17$ and $5.62$ TPa, respectively. The ZPE difference between $Pbcm$ and $Pmc2_1$, on the other hand, is found to be practically negligible, resulting in a shift of the transition pressure higher by less than $50$ GPa. Note that these estimates neglect the impact of zero-point pressure, which given the small differences in ZPE between the structures should be relatively minor.


One of the most intriguing suggestions regarding high-pressure ice is its metallization \cite{H2O_metallization_Grimsditch-BookCh-1985}. Pursuant to this, the electronic density of states for each structure shown in Fig.\ \ref{fig:phase-diag} was investigated. Except for $Fddd$ and $Cmm2$, all of the structures eventually become metallic with increasing pressure. However, $Pmc2_1$ was found to be insulating over its entire range of stability as was $P2_1$ up to $4.7$ TPa. Furthermore, given that DFT typically underestimates band gaps, the actual insulator-to-metal transition in $P2_1$ likely occurs at a much higher pressure. Although, metallization in $C2/m$ occurs at a pressure much lower than the $P2_1 \rightarrow C2/m$ transformation (in fact, it was found to be metallic at all pressures considered). It can therefore concluded that this phase transformation at $5.62$ TPa also marks the insulator-to-metal transition in ice, which is a significantly higher pressure than the recent prediction of $1.55$ TPa \cite{H2O_Cmcm_Pbca_Militzer-PRL-2010}.


In conclusion, AIRSS was used to determine the ground-state and metastable structures of ice at high pressures. The predicted transformation sequence is ice X \cite{H2O_ice-X_Grimsditch-PRL-1984} $\rightarrow$ $Pbcm$ ($300$ -- $400$ GPa \cite{H2O_Pbcm_Parrinello-PRL-1996}) $\rightarrow$ $Pmc2_1$ ($800$ -- $950$ GPa) $\rightarrow$ $P2_1$ ($1.17$ TPa) $\rightarrow$ $C2/m$ ($5.62$ TPa), where transition pressures have been indicated in parenthesis. No additional competitive structures were found during the AIRSS, even at $5$ TPa. Thus, any phases existing beyond $C2/m$ do so at much higher pressures, which are clearly outside of the range of experimental and most astrophysical applications. The $P2_1$ $\rightarrow$ $C2/m$ transformation was demonstrated to mark the insulator-to-metal transition in ice, which is also beyond pressures found inside even many giant planets, such as Jupiter, and it can therefore be concluded that the ice components in them remain insulating. Along with the recent work elucidating the high-pressure high-temperature phase diagram of water \cite{H2O_EOS_Redmer-PRB-2009}, the results presented herein provide a comprehensive picture of the high-pressure phase diagram of $\text{H}_2\text{O}$. 

\begin{acknowledgments}
J.\ M.\ M.\ was supported by DOE DE-FC02-06ER25794 and DE-FG52-09NA29456. This research was also supported in part by the National Science Foundation through TeraGrid resources provided by NICS under grant number TG-MCA93S030.
\end{acknowledgments}


\end{document}